\documentclass{aa} 
\usepackage[varg]{txfonts} 
\usepackage{natbib} 
\usepackage[bookmarks]{hyperref}
\hypersetup{pdfborderstyle={/S/U/W 1}}

\bibpunct{(}{)}{;}{a}{}{,}

\newcommand{\unit}{\mathrm} 
\newcommand{\name}{\mathrm}

\begin{document}

\title{Solid H$_2$ in the interstellar medium}

\author{A. F\"uglistaler \and D. Pfenniger} 
\institute{Geneva Observatory, University of Geneva, Sauverny, Switzerland\\ \email{andreas.fueglistaler@unige.ch}} 
\date{Received August 2017 / Accepted 24 January 2018}

\abstract 
{ Condensation of H$_2$ in the interstellar medium (ISM) has long been seen as a possibility, either by deposition on
	dust grains or thanks to a phase transition combined with self-gravity. H$_2$ condensation might explain the observed
	low efficiency of star formation and might help to hide baryons in spiral galaxies. }
{ Our aim is to quantify the solid fraction of H$_2$ in the ISM due to a phase transition including self-gravity for
	different densities and temperatures in order to use the results in more complex simulations of the ISM as subgrid
	physics. }
{ We used molecular dynamics simulations of fluids at different temperatures and densities to study the formation of
	solids. Once the simulations reached a steady state, we calculated the solid mass fraction, energy increase, and
	timescales. By determining the power laws measured over several orders of magnitude, we extrapolated to lower densities
	the higher density fluids that can be simulated with current computers. }
{ The solid fraction and energy increase of fluids in a phase transition are above $0.1$ and do not follow a power law.
	Fluids out of a phase transition are still forming a small amount of solids due to chance encounters of molecules. The
	solid mass fraction and energy increase of these fluids are linearly dependent on density and can easily be
	extrapolated. The timescale is below one second, the condensation can be considered  instantaneous. }
{ The presence of solid H$_2$ grains has important dynamic implications on the ISM as they may be the building blocks
	for larger solid bodies when gravity is included. We provide the solid mass fraction, energy increase, and timescales
	for high density fluids and extrapolation laws for lower densities. }


\maketitle

\section{Introduction}

The possibility of solid H$_2$ in the interstellar medium (ISM) was first proposed by \citet{van_de_hulst_solid_1949}.
Most of the subsequent literature concentrates on solid H$_2$ deposited on grains
\citep[e.g.][]{wickramasinghe_accretion_1968,hoyle_solid_1968,sandford_h2_1993}. \citet{walker_snowflakes_2013} analyses
the lifetime of solid H$_2$ grains of different sizes and concludes that they are longer than usually assumed. This
article considers pure condensed H$_2$, formed during a phase transition.

As is well-known in statistical physics, when a medium is in phase transition it  typically develops self-similar, power
law fluctuations at all scales. In the astrophysical context this means that self-gravity (large-scale) may interfere
with a phase transition (small-scale). In \citet[][hereafter FP2015 and FP2016]{fuglistaler_substellar_2015,
	fuglistaler_formation_2016}, and briefly recalled in Sect.\ 2, we showed that indeed a uniform fluid in phase transition
is also automatically gravitationally unstable because an overdensity fluctuation does not lead to a pressure increase,
but to a condensed phase increase, so nothing prevents gravity amplifying the overdensity. We showed with numerical
simulations that solid H$_2$ (and gaseous He) bodies can form in the ISM: a fluid in a phase transition first forms
small accumulations of molecules called oligomers (indeed, even out of a phase transition a very small percentage of
oligomers form). Thanks to their more important gravitational pull and dynamical friction, these oligomers attract each
other, segregating towards larger bodies. In the present work, we quantify these findings, giving scaling laws of the
solid H$_2$ fraction and the energy increase during its formation for different densities and temperatures to be used as
subgrid physics in larger simulations.

H$_2$ condensation properties are well-known from laboratory data \citep[e.g.][]{air_liquide_gas_1976}. On the other
hand, observations of solid H$_2$ is a different story altogether, as H$_2$ is only emitting at temperatures $\geq
512\,\unit{K}$. Molecular clouds are therefore mostly inferred from CO emissions \citep{bolatto_co--h_2013}, but it is
now well-established \citep{grenier_unveiling_2005,planck_collaboration_planck_2011} that some H$_2$ is not traced by
CO, the so-called `dark gas'.  There are other direct and indirect ways to detect H$_2$, as described in
\citet{combes_perspectives_1997}, but most of them only apply for gaseous H$_2$. \citet{lin_interstellar_2011} argue
that H$^+_6$ detection would indicate solid H$_2$, as this ion is not formed by gas-phase reactions.

Solid H$_2$ bodies, even planet-sized, are presently too small for direct detection, but ongoing surveys such as Gaia
might probe short microlensing events that were not detectable with previous microlensing surveys (Laurent Eyer, private
communication). We can speculate that dirty solid H$_2$ bodies could form in the coldest conditions in the ISM, such as
the densest parts of well self-shielded molecular clouds or in globulettes \citep{gahm_globulettes_2007,gahm_mass_2013}.
As soon as radiation and cosmic rays from stars  start to heat up these regions, volatile species such as H$_2$ and He
will be the first to evaporate, leaving bodies similar to asteroids or comets. So some interstellar comets, asteroids
\citep{meech_brief_2017}, or meteorites \citep{belyanin_petrography_2018} may be the remnants of former much larger
H$_2$ bodies containing traces of heavier atoms and molecules. The lifetime of solid H$_2$ km-sized bodies is much
longer than grains because cosmic rays and radiation cannot penetrate deep inside a solid body. The lifetime of km-sized
bodies can be evaluated by knowing the heating flux from stellar radiation and cosmic rays ($\sim
10^{-5}\,\unit{J/s/m^2}$) matching the H$_2$ latent heat ($\sim 500\, \unit{kJ/kg}$) in a body of mass $M$ and solid
H$_2$ density ($\sim 90\,\unit{kg/m^3}$). Lifetimes over a broad interval are obtained, such as $10^7 -
10^{10}\,\unit{yr}$, depending on the heating conditions and sizes of the assumed bodies.

The presence of solid H$_2$ in the ISM can have important implications. The formation of comet- or planet-sized solid
H$_2$ bodies may help to explain the observed low efficiency of star formation
\citep{mckee_theory_2007,draine_physics_2011,kennicutt_star_2012} if a large fraction of molecular cloud cores (the low
mass tail of the Salpeter law) condenses first as small bodies and then subsequently most of them evaporate as the
freshly formed stars heat them. The surviving substellar solid H$_2$ bodies might help to `hide' baryons for several
Gyr. Cold, dense H$_2$ clouds could then be a component of the dark matter in the outer part of galaxies
\citep[e.g.][]{pfenniger_is_1994-1, gerhard_baryonic_1996, revaz_simulations_2009}. \citet{pfenniger_is_1994} argue that
condensed H$_2$ may exist in the core of dense clouds, and \citet{wardle_thermal_1999} discuss that solid bodies would
help to thermally stabilize the clouds.

In addition to the formation of condensations due to H$_2$ condensation during a collapse in the ISM, the topic is more
general and may be relevant in different contexts where a phase transition acts together with gravity to produce
interesting physics, such as planetesimal formation in protoplanetary disks where temperatures can drop below
$10\,\unit{K}$ \citep{guilloteau_shadow_2016}.

\section{Physics}

In this section, we briefly summarize the findings on phase transition fluids described in FP2015/16.

\subsection{Equation of state}
\label{sec:LJ}

Usually, the ISM is described as an ideal gas.  However, an ideal gas cannot be in a phase transition. One of the
simplest equations of state to describe a phase transition is the van der Waals equation of state
\citep{van_der_waals_remarks_1910}. It reads in reduced units
\begin{equation}
\label{eq:vdW}
        P_\name{r} = \frac{8T_\name{r}}{\frac{3}{n_\name{r}}-1} - 3n_\name{r}^2 \ ,
\end{equation}
where $P_\name{r} = P / P_\name{c}$ is the reduced pressure, $n_\name{r} = \rho_\name{r} = n / n_\name{c}$ the reduced
number density, and $T_\name{r} = T / T_\name{c}$ the reduced temperature. The critical values of H$_2$ are $T_\name{c}
= 33.1\,\unit{K}$, $n_\name{c} = 9.34\cdot 10^{27}\,\unit{m}^{-3}$, and $P_\name{c}=1.30\cdot 10^6\,\unit{Pa}$
\citep{air_liquide_gas_1976}.

\begin{figure}[t] 
        \centering
        \includegraphics[width=8.8cm]{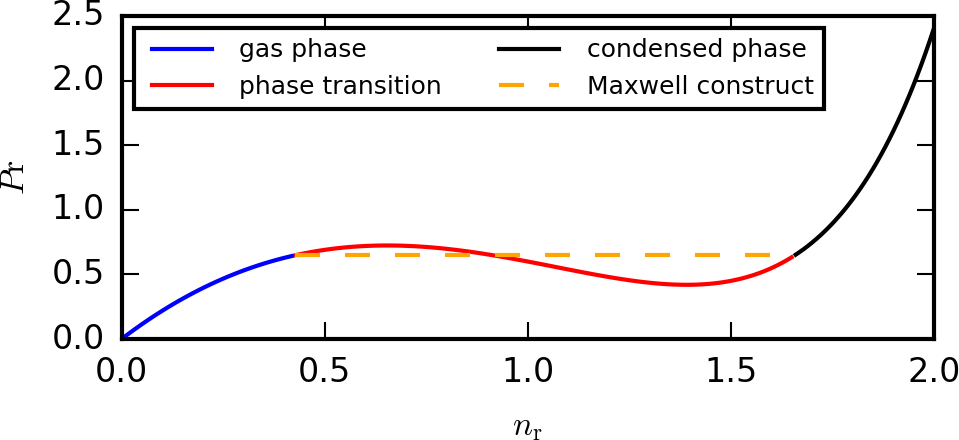}
        \caption{van der Waals phase diagram for a fluid with $T_\name{r}=0.9$. The gas phase and condensed phase are linked by
        the phase transition. As a negative $\left({\partial P / \partial\rho}\right)$ is unphysical, this part of the
        equation is replaced by the Maxwell construct.}
        \label{fig:vdw2d}
\end{figure}

Figure \ref{fig:vdw2d} shows the phase diagram for a van der Waals fluid. A fluid  with $T < T_\name{c}$ can be in a
gaseous or condensed phase. When the fluid is in a phase transition, i.e. when the two phases coexist, a constant
pressure marked by a horizontal line replaces the van der Waals equation of state.  This constant pressure level is
determined by the Maxwell equal area construct, demanding a total zero $P\cdot v = P/n$ work for an adiabatic cycle
between the fully gaseous to the fully condensed state \citep{clerk-maxwell_dynamical_1875,johnston_thermodynamic_2014}.
The isothermal density derivative of such a fluid is
\begin{equation}
\label{eq:dvdW}
\left(\partial P \over \partial \rho\right)_T  = 
\begin{cases}
\quad 0 & \quad \text{on the Maxwell line} \\
\quad {24 T_r \over m (n_r-3)^2} - {6n_r\over m} & \quad \text{else} \\
\end{cases} \ .
\end{equation}
The van der Waals equation of state describes a fluid from a continuum point of view, which is unable to encompass the
rich physics taking place in a phase transition, such as phase separation in a gravity field (rain). In order to include
these effects we use a particle-based code where particles have a typical molecular interaction potential. This
potential, the Lennard-Jones potential, is able to reproduce in the average the van de Waals equation of state, and
automatically takes care of the correct latent heat of the H$_2$ molecules, and the Maxwell construct
\begin{equation}
\Phi_{\name{LJ}}(r) = 4{\epsilon \over m}\left[\left({\sigma \over r}\right)^{12} - \left({\sigma \over r}\right)^{6}\right] \ ,
\end{equation}
where $m$ is the mass of a molecule. The Lennard-Jones energy $\epsilon$ and distance $\sigma$ can be linked to the van
der Waals equation of state with the following relations \citep{caillol_critical-point_1998}:
\begin{eqnarray}
\label{equ:Tc}
T_\name{c} &=& 1.326 \, {\epsilon \over k_\name{B}} \ , \\
\label{equ:nc}
n_\name{c} &=& 0.316 \, \sigma^{-3} \ , \\
\label{equ:Pc}
P_\name{c}  &=& 0.157\, { \epsilon \over k_\name{B}\sigma^{3} } \ .
\end{eqnarray}
The critical pressure is calculated using the critical compression factor $Z_\name{c}=p_\name{c}/(T_\name{c}
n_\name{c})=3/8$ \citep{johnston_thermodynamic_2014}.

\subsection{Gravitational stability of fluids in a phase transition}

The Jeans criterion \citep{jeans_stability_1902} states that a fluid is gravitationally unstable if a perturbation has a
wavelength larger than
\begin{equation} 
\lambda > 
\lambda_\name{J} \equiv \sqrt{\left({\partial P \over \partial\rho}\right)_s\frac{\pi}{G\rho}} \ , 
\label{eq:Jeans} 
\end{equation}
where $\left({\partial P / \partial\rho}\right)_s$ is the speed of sound $c_\name{s}$ squared, $G$ the gravitational
constant, $P$ the pressure, $\rho$ the density, and $s$ the entropy.

In the case of a perfect gas, $c_\name{s}^2 = \gamma k_\name{B} T/m$, with $\gamma$ the adiabatic index, $k_\name{B}$
the Boltzmann constant, $T$ the temperature, and $m$ the molecular weight.  So for a perfect gas $\lambda_\name{J}$ is
always positive; that is, there is always a scale below which the gas is gravitationally stable.

However,  $\left(\partial P / \partial \rho\right)_s = \gamma \left(\partial P / \partial \rho\right)_T$ with the
adiabatic index $\gamma > 0$. Therefore, a fluid in a phase transition has $({\partial P / \partial\rho})_s =
c_\name{s}^2 = 0$ (Maxwell's construct), consequently $\lambda_\name{J} = 0$. Formally, a fluid in a phase transition is
gravitationally unstable even at microscopic scale.

\subsection{Formation of condensations}

The Jeans length (Eq. (\ref{eq:Jeans})) of an ideal gas is
\begin{equation} 
\lambda_\name{J,id} = \sqrt{\frac{\pi \gamma k_\name{B}T}{G\rho m}} \ . \label{equ:Jeansideal}
\end{equation}
We showed (FP2015 and FP2016) that a fluid with $\lambda > \lambda_\name{J,id}$  collapses in a short time (Jeans time),
heats up, and forms a gaseous body. This happens independently whether the fluid is in a pure gaseous phase or in a
phase transition. On the other hand, if $\lambda < \lambda_\name{J,id}$ we need to distinguish between a pure gaseous
fluid and a fluid in a phase transition. The former will remain gaseous, the perturbation being simply a sound wave.

Fluids in a phase transition, however, are always gravitationally unstable (see above). Small condensations quickly form
due to the phase transition. After a short period a new equilibrium is reached and the fluid has two phases, gaseous and
solid. After that, the condensation fraction remains the same, but the grain size increases.

At first, the condensations only consist of a few molecules called oligomers. Thanks to dynamical friction, these
oligomers attract each other and form bigger clumps, and the condensation size increases from snowballs, then comets,
and  up to planetoids. The final size of the solid body depends on the total mass, the solid mass fraction, and the
formation timescale.  The latter two are dependent on the density (see \S \ref{sec:Results}).

The accumulation of molecules towards oligomers and then larger bodies happens first by three-body interactions where
one of the molecules carries away the excess energy. This leads to a temperature increase of the gaseous part of the
fluid. However, this energy increase is only a fraction of the initial temperature (see Fig. \ref{fig:results} (B) and
FP2015, FP2016), and gets smaller with increasing initial temperature. Therefore, the temperature does not exceed the
critical value of H$_2$. If there is rapid cooling, the fluid remains at its initial temperature.

A non-trivial result of our previous works is that the solid mass fraction $M_\name{solid}/M_\name{tot}$ is the same
with or without gravity for phase transition fluids. The formation of larger bodies happens only by accumulating solid
objects, and thus does not increase the solid fraction. In order to include phase transition effects in astrophysical
hydrodynamical simulations it is therefore crucial to know this value for different densities and temperatures.

\subsection{Paths to a phase transition state}

\begin{figure}[t] 
        \resizebox{\hsize}{!}{\includegraphics{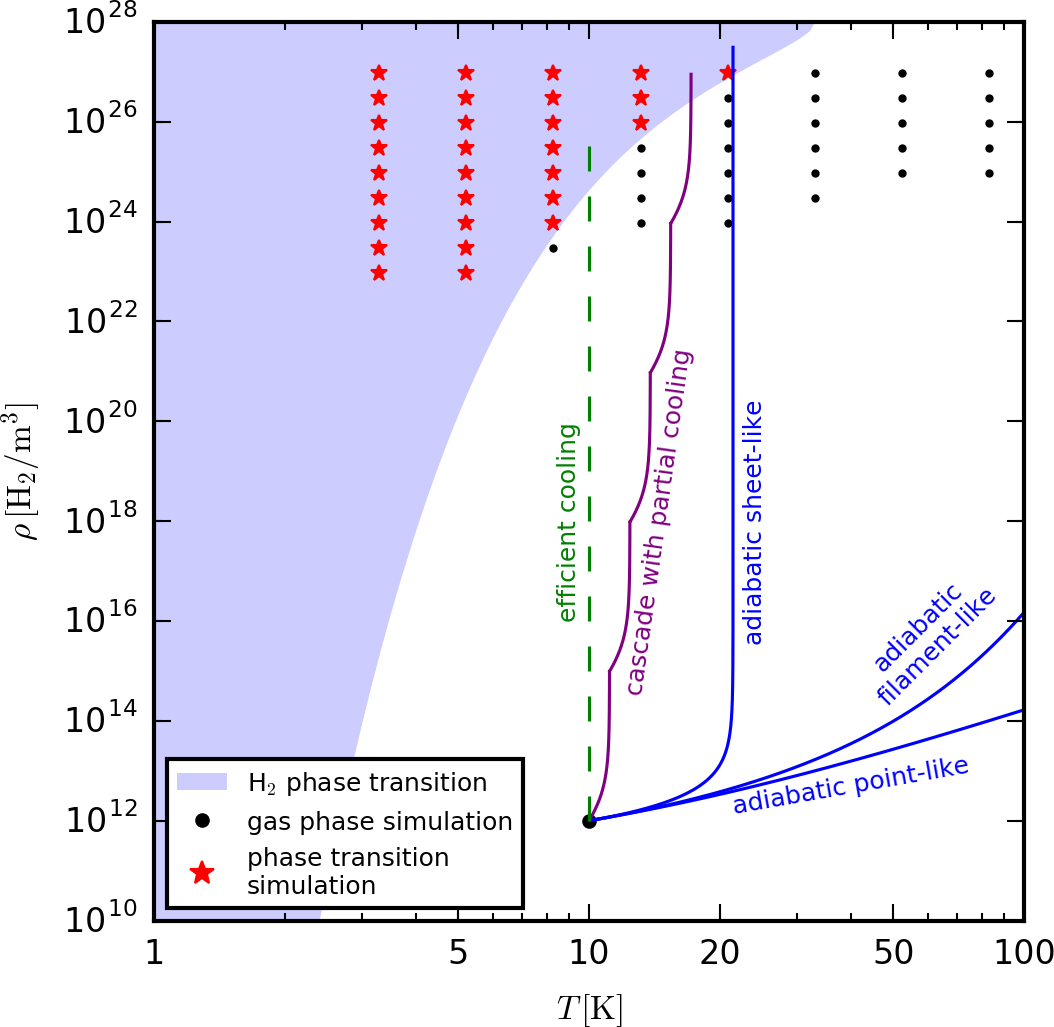}} 
        \caption{Phase diagram of H$_2$. Different adiabatic collapsing geometries and collapses including cooling are
	shown of a fluid with typical properties of a molecular cloud: $T=10\,\unit{K}$, $n=10^{12}\,\unit{m}^{-3}$. }
        \label{fig:T-rho} 
\end{figure}

The H$_2$ phase transition happens well at typical temperatures ($<33\,\unit{K}$) for the cold ISM, but at very high
densities ($>10^{20}\unit{m}^{-3}$) more typical of laboratory conditions. Thus, solid H$_2$ is usually not expected to
form directly in traditional collapse scenarios (one-step spherical collapse). However, two ingredients must be
considered. Typical gravitational collapses occur along a sequence of geometries: first pancake (2D), then filament
(1D), and finally point (0D) collapses \citep{lin_gravitational_1965,zeldovich_gravitational_1970}, and this in a
recursive, fractal manner until the microphysics changes and reacts to the increased density and pressure. In FP2016, we
discuss different collapsing geometries. A pancake or sheet-like collapse is not only the fastest collapsing geometry,
but also leads at most to a doubling of temperature when density increases by more than a factor 100 in adiabatic
conditions.

Cooling by radiation loss, depending on the opacity of the medium, leads to a smaller or negligible temperature
increase. In addition, in sheet-like geometry the opacity increases only slightly, so reaching an optically thick
condition is harder. Taking into account the multiscale, recursive nature of gravitational collapses, a cascade of
collapses at higher and higher densities should proceed, allowing  phase-transition conditions to be reached. Thus, a
cascade of sheet-like, transparent to radiation collapses, combined with some cooling, can lead to a H$_2$ phase
transition. Along the cascade of collapses the specific angular momentum present in the initial density fluctuations
must increase, increasingly favouring (as the scale decreases) the disk geometry instead of mere non-rotating pancake
geometry.

Figure \ref{fig:T-rho} shows the phase diagram of H$_2$. A gas with typical properties of molecular clouds is compressed
using different paths. While one-step filament- and point-like collapses lead to divergent temperatures, a sheet-like
collapse without cooling only doubles the temperature. Assuming fast cooling, an increase of density without any
temperature increase is conceivable, as the fluid opacity is not increased much during a sheet-like collapse. We also
sketch the case of a cascade including partial cooling, which reaches a H$_2$ phase transition after five recursive
steps.

\section{Simulations}

Our goal is to calculate the solid mass fraction $M_\name{solid}/M_\name{tot}$, the energy increase $\Delta
E_\name{cond}$, and the timescale $\tau_\name{cond}$ of the formation of oligomers due to a phase transition. Figure
\ref{fig:T-rho} indicates the different temperature and densities of the simulations. The lower density limit is given
by the needs of computing power: with decreasing density molecular encounters become rarer and the fraction
$M_\name{solid}/M_\name{tot}$ lower. As a result, the time needed to reach a stable value for a fraction is much longer.
In addition, to have a high enough resolution, more particles are needed. However, as we  see in \S \ref{sec:Results},
we do reach sufficiently low densities in order to obtain a power law regime that allows us to extrapolate to lower
densities with good confidence.

As described in more detail in FP2015 and FP2016, we use the popular molecular dynamics simulator LAMMPS
\citep{plimpton_fast_1995} to perform our simulations using the Lennard-Jones potential. As we want to calculate the
fraction $M_\name{solid}/M_\name{tot}$, which is independent of the strength of the gravitational potential, we do not
need  to include gravity in these simulations.

Most of the simulations have $N=10^{5}$ molecules (see \S \ref{sec:Scaling} for scaling), distributed initially randomly
in a cube with periodic boundary conditions. We choose the volume $V$ of the cube for different $n = N/V$ with number
densities of $\log_{10}(n_\name{r})=-6, -5.5, \ldots, -1$. The molecules have a Maxwellian velocity distribution with
temperatures of $\log_{10}(T_\name{r})=-1, -0.8, \ldots, 0.4$. The cut-off distance is $r_\name{c} = 4 \sigma$.

\begin{figure}[t] 
    \resizebox{\hsize}{!}{\includegraphics{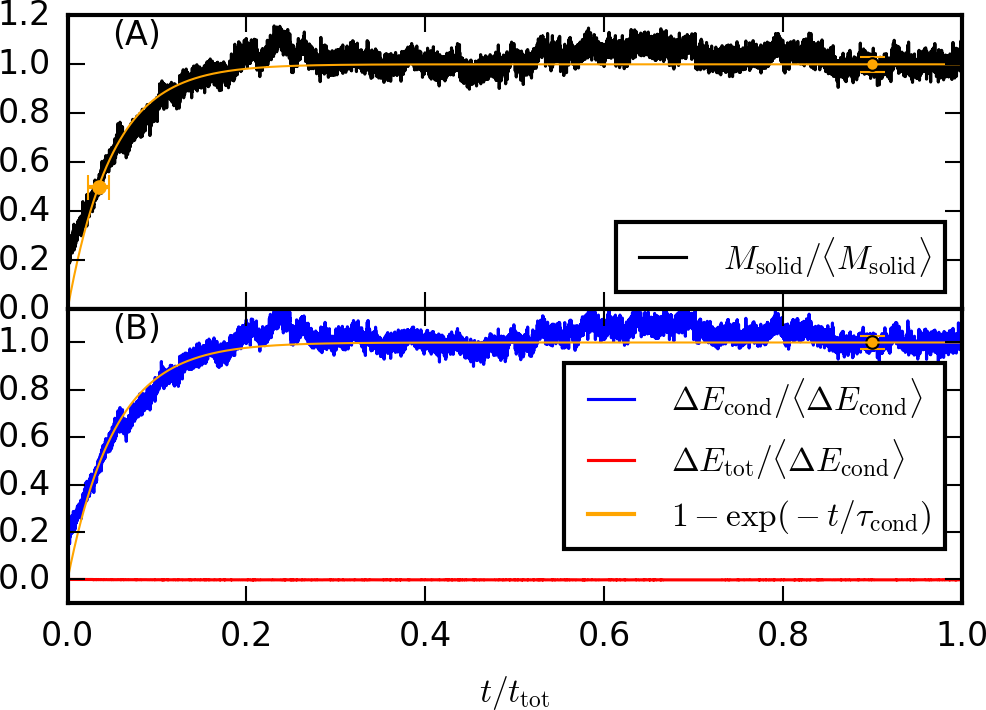}} 
    \caption{Evolution of the solid mass fraction $M_\name{solid}$, kinetic energy increase $\Delta E_\name{cond}$, and total
        energy increase $\Delta E_\name{tot}$ of the simulation with $n_\name{r} = 10^{-3.5}$, $T_\name{r} = 10^{-0.6}$, and
        $N = 10^5$. $\langle M_\name{solid}\rangle$ and $\langle \Delta E_\name{cond}\rangle$ are the average values and the
        error bars show the standard deviation. The extrapolation is $\propto 1-\exp(-t/\tau)$ with timescale $\tau$. }
    \label{fig:n-35T-06}
\end{figure}

\section{Results}
\label{sec:Results}

Figure \ref{fig:n-35T-06} shows a typical evolution of the solid mass fraction and kinetic energy increase during an
adiabatic simulation. Both values follow an exponential law $\propto 1 - \exp(-t/\tau)$, with time $t$ and timescale
$\tau$. Once the simulation reaches a steady-state value, we measure the solid mass fraction and energy increase. These
average values, $\langle M_\name{solid}\rangle$ and $\langle\Delta E_\name{cond}\rangle$, and the standard deviation are
shown in Fig. \ref{fig:results} (A) and (B).

At high density and  low temperature (especially $\log_{10}(T_\name{r}) = -1$ and $-0.8$), there is a high fraction of
condensations. These are the fluids which are `officially', according to the van der Waals equation of state, in a phase
transition. However, even fluids out of a phase transition, which have lower densities and/or higher temperatures, form
a small amount of condensations.

The amount of condensations in phase transition fluids cannot be linked linearly to the density or temperature, as the
intermolecular interactions are  non-linear ($\propto r^{-6}$ and $r^{-12}$). The fluids out of a phase transition form
condensations due to chance interaction between two or more molecules. The number of these interactions is linearly
dependent on the density, and the probability of forming condensations increases with decreasing temperature. They
follow a linear law:
\begin{equation}
\label{equ:Msolid}
\left({M_\name{solid} \over M_\name{tot}}\right)_T \propto n \ .
\end{equation}
The simulations are adiabatic: the condensation formation implies a decrease in potential energy and therefore an
increase in kinetic energy. One can recognize the similarities between the energy increase (Figure \ref{fig:results}
(B)) with the solid mass fraction (A): again the phase transition fluids are non-linear, but for fluids with low
densities and/or high temperatures the energy of condensation follows a linear law
\begin{equation}
\label{equ:Ekin}
\left({\Delta E_\name{cond} \over E_\name{kin0}}\right)_T \propto   n  \ ,
\end{equation}
where $E_\name{kin0}$ is the initial kinetic energy.

For all practical astrophysical uses, condensation should be considered an instantaneous process. Even simulations with
many time steps (the simulations had up to $10^9$ time steps) are orders of magnitude shorter than $1\unit{s}$ in real
time. We measure the condensation timescale by identifying the first appearance of $\langle M_\name{solid}\rangle$, and
we average $\tau$ for all values up to that point. As seen in Fig. \ref{fig:n-35T-06}, the resulting value approximates
the data rather well.

Figure \ref{fig:results} (C) shows the condensation timescales of the simulation for reaching the average $\langle
M_\name{solid}\rangle$ value. No universal power law seems to govern the timescales. The timescales of low temperature
fluids ($T_\name{r}\leq 10^{-0.4}$) follow an inverse power law:
\begin{equation}
\label{equ:tcond}
\left(\tau_\name{cond}\right)_T \propto n^{-1} \ .
\end{equation}
The timescales of high temperature fluids ($T\geq 10^{-0.2} T_\name{c}$) follow an inverse square root law:
\begin{equation}
\label{equ:tcond1}
\left(\tau_\name{cond}\right)_T \propto n^{-1/2} \ .
\end{equation}

\subsection{Scaling}
\label{sec:Scaling}

\begin{figure}[t] 
    \resizebox{\hsize}{!}{\includegraphics{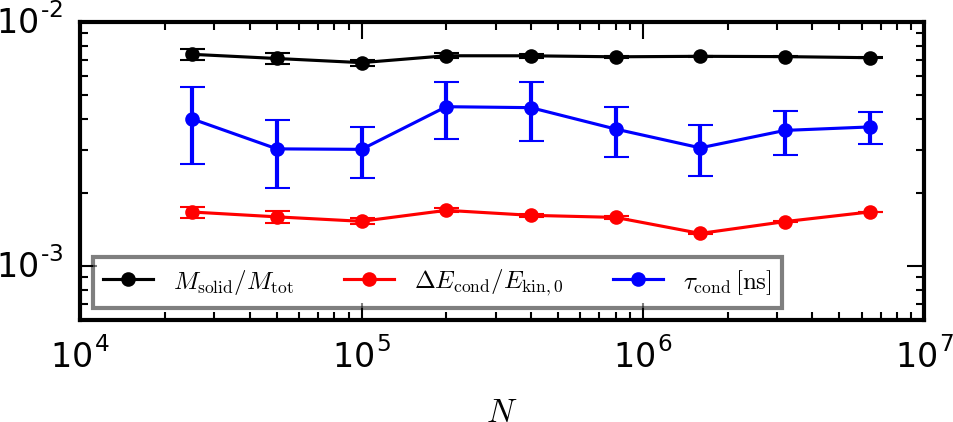}} 
    \caption{Scaling of the simulation with $n_\name{r} = 10^{-3.5}$, $T_\name{r} = 10^{-0.6}$, and different numbers of
        molecules N. Average values and standard deviation are shown for the solid fraction $M_\name{solid}/M_\name{tot}$,
        condensation energy $\Delta E_\name{cond}$, and condensation timescale $\tau_\name{cond}$.}
    \label{fig:scale}
\end{figure}

In order to show the independence from the number of particles $N$, we ran several simulations with different values for
$N$. Figure \ref{fig:scale} shows the solid mass fraction, energy increase, and timescale as a function of $N$ for such
a simulation. All three values remain constant.

\subsection{Extrapolation}
\label{sec:Extrapolation}

The values of $M_\name{solid}/ M_\name{tot}$, $\Delta E_\name{cond}$, and $\tau_\name{cond}$ are non-linear for fluids
in a phase transition. However, they can be extrapolated for fluids out of a phase transition. Fluids with temperatures
of $T_\name{r} = 0.1$ are in a phase transition for densities down to ISM values (however, densities below
$10^{-6}n_\name{c}$ cannot be simulated using molecular dynamics on today's computers). Extrapolation laws are given for
fluids with temperatures $T_\name{r} > 0.1$.

Least-squares extrapolations of the type $a \log_{10}(n) + b$ for $M_\name{solid}/ M_\name{tot}$ and $\Delta
E_\name{cond}$ show that  $a = 1 \pm 0.05\, \forall T$. A linear approach is thus appropriate for these two laws. The
least-square extrapolation for the condensation timescale $\tau_\name{cond}$ gives higher errors with $a = -1 \pm 0.2$
for $T_\name{r} \leq 10^{-0.4}$, and $a = -0.5 \pm 0.2$ for $T_\name{r} \geq 10^{-0.2}$.

We give extrapolation laws for the timescales, but caution  that they are speculative. This is  not a large practical
issue, however,  as the linear laws overestimate the timescales, and even the extrapolated values for densities as low
as the ISM are below one year.

The extrapolation laws read:
\begin{eqnarray}
\label{equ:Msolidlsq}
\log_{10}\left({M_\name{solid} \over M_\name{tot}}\right) &=& \log_{10}(n_\name{r}) + M(T_\name{r}) \ , \\
\label{equ:Ekinlsq}
\log_{10}\left({\Delta E_\name{cond} \over E_\name{kin0}}\right) &=& \log_{10}(n_\name{r}) + E(T_\name{r}) \ , \\
\label{equ:tcondlsq}
\log_{10}\left({\tau_\name{cond}\over 1\unit{s}}\right) &=& \begin{cases}
-\log_{10}(n_\name{r}) + \Theta(T_\name{r}) & \text{for } T_\name{r} \leq 10^{-0.4} \\
-\log_{10}(n_\name{r})/2 + \Theta(T_\name{r}) & \text{for } T_\name{r} \geq 10^{-0.2} 
\end{cases} \ .
\end{eqnarray}
The values for $M(T_\name{r})$, $E(T_\name{r}),$ and $\Theta(T_\name{r})$ are given in Table \ref{tab:lstsq}.

\begin{table}[ht] 
    \caption{Constants for Eqs. (\ref{equ:Msolidlsq}) -- (\ref{equ:tcondlsq}).} 
    \label{tab:lstsq} 
    \centering 
    \begin{tabular}{r | l l l } 
        \hline\hline 
        $\log_{10}(T_\name{r})$  & $M(T_\name{r})$ & $E(T_\name{r})$ & $\Theta(T_\name{r})$  \rule{0pt}{2.6ex}\\ 
          \hline  
              $-0.8$ & $2.0169$ & $1.3738$ & $-11.4190$\\
              $-0.6$ & $1.3512$ & $0.7036$ & $-11.9727$\\
              $-0.4$ & $0.9420$ & $0.2477$ & $-12.3988$\\
              $-0.2$ & $0.7101$ & $-0.0715$ & $-11.5131$\\
              $-0.0$ & $0.5732$ & $-0.3301$ & $-11.6386$\\
              $0.2$ & $0.4822$ & $-0.5636$ & $-12.0104$\\
              $0.4$ & $0.4363$ & $-0.8622$ & $-12.1677$\\        
          \hline 
    \end{tabular} 
\end{table}

\section{Conclusions}

In astrophysical conditions, solid H$_2$ formation can be seen as an instantaneous process with respect to gravitational
processes. The solid mass fraction $M_\name{solid}/M_\name{tot}$ is high for fluids in a phase transition, ranging from
$0.01$ to $>0.1$. However, it does not drop to $0$ for fluids out of a phase transition. A small fraction of the gas is
always condensing as a result of the encounters of two or more molecules. The solid fraction is small and declines
linearly with decreasing density. Nonetheless, even a mass fraction as small as $10^{-6}$ can make a difference, e.g. in
planet formation where the ratio $M_\name{planet}/M_\odot$ lies in the same range.

We present extrapolation laws to calculate the solid mass fraction $M_\name{solid}/M_\name{tot}$ and amount of energy
increase $\Delta E_\name{cond}$ due to the formation of oligomers for fluids out of a phase transition for temperatures
from $T = 0.15\,T_\name{c}$ to $2.5\,T_\name{c}$. The extrapolated values, together with the non-linear values for
fluids in a phase transition from the simulations can be interpolated for subgrid physics in larger, continuous fluid
simulations.

A phase transition in the ISM, especially the H$_2$ transition in view of its high abundance, may have important
dynamical implications which cannot be described using continuous physics, as is done in current simulation codes. 
Since in a phase transition all scales interact, and such a medium is always gravitationally unstable, the combination
of gravity and phase transition is ideal to form bodies with sizes ranging from small oligomers to snowballs up to
comet-sized bodies. Oligomers are seeds that facilitate the condensation of larger bodies on  small scales, while
gravitational condensations (clumps, spiral arms) on large scales combine with the phase transition scales to form
bodies dominated by gravity (spherical bodies).

The ideal locations for condensing H$_2$ in comet- or planet-sized bodies are cold starless cores, self-shielded from
radiation in molecular clouds in the outskirts of galaxy disks where the local temperature should drop well below
$10\,\unit{K}$. Possibly bodies mostly made of H$_2$ may form there and subsequently become a collisionless component of
the disk until heating evaporates them, leaving heavier element bodies similar to comets.

\begin{acknowledgements}
This work is supported by the STARFORM Sinergia Project funded by the Swiss National Science Foundation.  We thank the
LAMMPS team \footnote{\texttt{https://lammps.sandia.gov}} for providing a powerful open source tool to the scientific
community. We thank the referee for a thorough reading of the manuscript and constructive comments which substantially
improved the paper.
\end{acknowledgements}

\bibliographystyle{aa} 
\bibliography{SolidISM} 
\addcontentsline{toc}{section}{\refname}

\appendix

\renewcommand{\thefigure}{A\arabic{figure}}

\setcounter{figure}{0}

\begin{figure}[p] 
    \resizebox{\hsize}{!}{\includegraphics{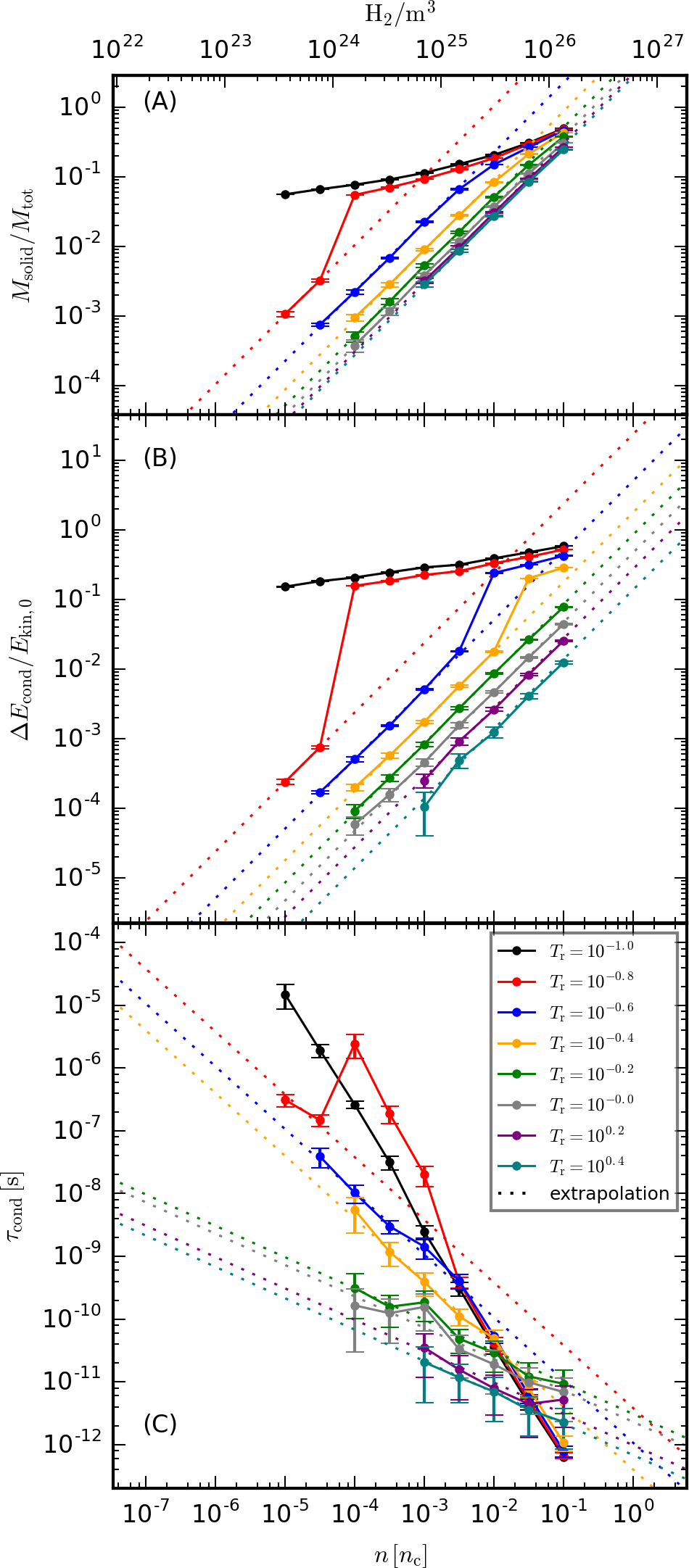}} 
    \caption[]{Simulation results for different temperatures $T_\name{r}$ and number densities $n_\name{r}$. The points
        show the average values, and the error bars show the standard deviation.\\
        (A) Solid mass fraction.\\
        (B) Temperature increase during condensation.\\
        (C) Condensation time.}
    \label{fig:results} 
\end{figure}

\end{document}